\documentclass[%
 amsmath,amssymb,
 aps,
prc,
]{revtex4-2}

\usepackage{xcolor}
\usepackage{hyperref}
\usepackage{graphicx}
\usepackage{dcolumn}
\usepackage{bm}

\begin{document}


\title{Decay properties of the $d^*(2380)$ hexaquark multiplet}
\author{M. Bashkanov}
\email{mikhail.bashkanov@york.ac.uk}
 \affiliation{Department of Physics, University of York, Heslington, York, Y010 5DD, UK}
\author{G. Clash}%
\affiliation{Department of Physics, University of York, Heslington, York, Y010 5DD, UK}
\author{M. Mocanu}%
\affiliation{Department of Physics, University of York, Heslington, York, Y010 5DD, UK}
\author{M. Nicol}%
\affiliation{Department of Physics, University of York, Heslington, York, Y010 5DD, UK}
 
\author{D.P. Watts}%
\affiliation{Department of Physics, University of York, Heslington, York, Y010 5DD, UK}


\date{\today}

\begin{abstract}
The recently discovered $d^*(2380)$ hexaquark is expected to be the lightest member of  an extended SU(3) antidecuplet of hexaquark states. The experimental search for the other heavier and strange partners of the $d^*(2380)$ in the antidecuplet is a challenging task. Evaluating the most appropriate methodologies necessitates some understanding of the underlying properties of the decuplet states such as mass, width and decay branches. In this paper we provide estimates of these key properties for all decuplet states, extrapolating from information and insights garnered for the $d^*(2380)$. The predictions form a basis for the design of future discovery experiments. 
\end{abstract}

\maketitle


\section{\label{sec:Intro} Introduction}

The properties of the six-quark containing $d^*(2380)$ (hexaquark) have been established quite rigorously in recent years following its observation in proton-neutron scattering and pionic fusion reactions~\cite{mb,MB,MBC,TS1,TS2,MBA,MBE1,MBE2,BCS}. It has a mass of $M_{d^*}=2380$~MeV, vacuum width $\Gamma = 70$~MeV and quantum numbers $I(J^P)=0(3^+)$. All of the strong decay branches have been identified and measured in experiment~\cite{BCS}. The electromagnetic properties of the $d^*(2380)$ were also investigated recently from measurements of its photoexcitation from deuteron targets~\cite{mbMainz, MBPy, BWP20}, with further programmes planned~\cite{MAMI_proposal}. The particle has also recently been shown to have potential impact in astrophysics\cite{nstars,nstars1,Cond}.

The search for a strange SU(3) partners of the $d^*(2380)$ is a natural next step for hexaquark studies. Characterisation of additional members of the antidecuplet would provide valuable insights into the underlying physics of hexaquark systems. However, as there is an abundance of states and associated decay channels from which to focus the initial searches, some prioritisation needs to take place. In this paper we develop a simple model to identify the most appropriate final state reaction channels for all members of the antidecuplet, including estimates of the partial decay widths.

\section{$d^*(2380)$ multiplet}
\begin{figure}[!h]
\begin{center}
        \includegraphics[width=0.4\textwidth,angle=0]{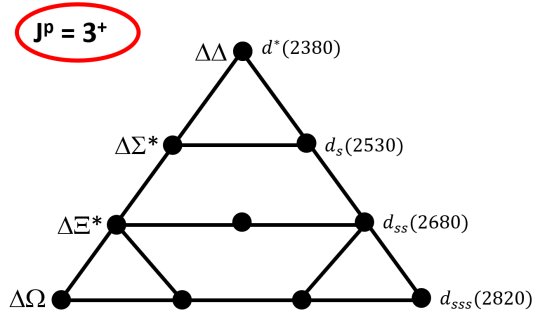}
\end{center}
\caption{$d^*(2380)$ multiplet}
\label{mult}
\end{figure}

From the unitary group theory of the strong interaction for the light quark (u,d,s) sector any strongly interacting particle, such as the  $d^*(2380)$, should be part of an SU(3) multiplet. For the case of the $d^*(2380)$ it would be expected to be a member of an antidecuplet, Fig.~\ref{mult}. The spectroscopic study of the other multiplet members would provide important new constraints on the $d^*$ internal structure, complimentary to that achievable in Form-Factor studies. In a simple molecular picture, the $d^*$ SU(3) multiplet would derive from the coupling of two baryon decuplet members bound by long-range pion exchange: corresponding to $\Delta\Delta$ for the $d^*(2380)$ to $\Delta\Omega$ for the $d_{sss}$. However, since the pion does not have a coupling to strange quarks, there is an expectation in such molecular systems that the binding energy should decrease with increasing strangeness content of the state. Conversely in a compact hexaquark (non-molecular) picture the binding energy should {\em increase} with increasing strangeness content
as the presence of heavier s-quarks would imply stronger binding.
Regardless of the inferred structure the $10^*$ antidecuplet decay properties are largely driven by the SU(3) symmetry and available phase space for the decay products. Being a $10^*$  SU(3) antidecuplet, the decay couplings are limited to either octet+octet, $8\oplus8$ or decuplet+decuplet, $10\oplus10$ baryons. The possible decay branches, deriving from the established $8\oplus8$ and $10\oplus10$ baryonic members members are summarised in table~\ref{Tab1}.
\begin{table}[ht]

\centering \protect\caption{Expected decay branches of the $d^*(2380)$ SU(3) multiplet}
\vspace{2mm}
{%
\begin{tabular}{|l|c|c|}
\hline
Particle  &  $8\oplus8$ & $10\oplus10$\\
\hline
$d^*$    & $pn$  & $\Delta\Delta$ \\
$d_s$    & $\frac{1}{\sqrt{2}}(N\Lambda-N\Sigma)$  & $\Delta\Sigma^*$ \\
$d_{ss}$    & $\frac{1}{\sqrt{6}}(\sqrt{2}N\Xi-\Sigma\Sigma+\sqrt{3}\Sigma\Lambda)$  & $\frac{1}{\sqrt{3}}(\sqrt{2}\Delta\Xi^*+\Sigma^*\Sigma^*)$ \\
$d_{sss}$    & $\Sigma\Xi$  & $\frac{1}{\sqrt{2}}(\Delta\Omega-\Sigma^*\Xi^*)$ \\
\hline
\end{tabular}} \label{Tab1}
\end{table}

\section{Formalism}
In our calculations we assume that the Breit-Wigner width of all the resonances is energy dependent and that the coupling constants for the decays into $8\oplus8$ and $10\oplus10$ are independent of the decaying particle's hypercharge/strangeness. This enables the total width of the decaying state to be expressed as
\begin{equation}
\Gamma_{tot}=\Gamma_8+\Gamma_{10},
\end{equation}
with $\Gamma_8$ corresponding to the partial width for hexaquark decay into $8\oplus8$ and $\Gamma_{10}$ for $10\oplus10$.
In case of $8\oplus8$ decay the final state particles are stable against strong decay so the $\Gamma_8$ decay width can be expressed as 
\begin{equation}
    \Gamma_8=g_8^2p^{2L+1}F_{8}(p)
\end{equation}
with
\begin{equation}
    F_{8}(p)=\frac{R^{2L}}{1+R^{2L}p^{2L}}
\end{equation}
where $g_8$ is the coupling constant of the hexaquark decay into $8\oplus8$, $p$ is the momentum of the ejectile in the hexaquark rest frame, $L$ is the angular momentum in the system and $F(p)$ is a Form-Factor, usually introduced to account for potential barriers, similar to Ref.~\cite{PBC,Kuk2}. For a $J^p=3^+$ particle decaying into two $J^p=1/2^+$ there are two possible scenarios - a $^3D_3$ partial wave with $L=2$ and $^3G_3$ partial wave with $L=4$. From partial wave analysis of the $d^*(2380)\to pn$ reaction we know that the majority ($\sim 90\%$) of the $8\oplus8$ decay proceeds through the $^3D_3$~\cite{MBE1,MBE2,TSE1}. In our calculation we will assume that all $8\oplus8$ proceed with $L=2$. This assumption leads to minor corrections, but allows a significant reduction in the required number of coupling constants. Since all the $3^+$-hexaquarks lie far above the $8\oplus8$ threshold the $\Gamma_8$ width is expected to be nearly constant throughout the resonance.

The $10\oplus10$ channel is a lot more challenging since the baryon decuplet contains resonant states with rather sizeable width an associated strong energy dependence. The $\Gamma_{10}$ width is expected to vary significantly between different resonances. We have calculated the $\Gamma_{10}$ as follows

\begin{align}
        \Gamma_{10}=\gamma_{10}\int dm_1^2dm_2^2F(q_{10})|D_{D_{1}}(m_1^2)D_{D_{2}}(m_2^2)|^2 \\
        F(q_{10})=\frac{\Lambda^2}{\Lambda^2+q^2_{10}/4},\;\;\; \Lambda=0.16~GeV/c \\
        D_{D}=\frac{\sqrt{m_{D}\Gamma_{D}(q_{M})/q_{M}}}{M_{BM}^2-m_{D}+im_{D}\Gamma_{D}^{tot}(q_{M})} \\
        \Gamma_{D}=\gamma(q_{M})^3\frac{R^2}{1+R^2(q_M)^2} 
\end{align}

here $F(q_{10})$ is a Form-Factor which depends on the relative momentum ($q_{10}$) between the two decuplet baryons in the hexaquark decay. We follow the prescription of Ref.~\cite{PBC}, and parametrize this dependence in a monopole form with a cut-off parameter $\Lambda$~\cite{EndNote1}. $D_{D_1}/D_{D_2}$ are the propagators for the decuplet of baryons, with $m_1/m_2$ the baryon masses (or correspondingly the invariant mass of the meson-baryon system $M_{BM}$ from the decuplet baryon decay into the octet of mesons and the octet of baryons) and $m_{D}$ being their nominal Breit-Wigner masses. The energy dependent width of the baryon decuplet decay, $\Gamma_D$, is parametrised in a standard form taking a $P$-wave decay resonance with Blatt-Weisskopf barrier factors of $R=6.3$~GeV/c and $\gamma = 0.74, 0.4, 0.14$ for the $\Delta,\Sigma^*$ and $\Xi^*$ respectively~\cite{Gal2}. $\Gamma_D^{tot}$ is the sum of all partial widths. The width of the $\Omega$ is considered to be zero, since it is stable with respect to strong decays. The Form-Factor parameters are fixed based on the $d^*\to \Delta\Delta\to d\pi\pi$ invariant mass distributions of Ref.~\cite{MB}. The $g_8$ constant is fixed to the value extracted from $\Gamma(d^*\to pn) = 8$~MeV. The $\gamma_{10}$ is taken as a normalisation factor and is fixed to reproduce $\Gamma_{10}=\Gamma_{D_1}+\Gamma_{D_2}$ at zero binding energy, e.g $\Gamma(d^* \to \Delta\Delta) = 2\cdot \Gamma_{\Delta}$ for $M_{d^*}=2\cdot M_{\Delta}$.

\section{Results}

\begin{figure}[!h]
\begin{center}
        \includegraphics[width=0.30\textwidth,angle=0]{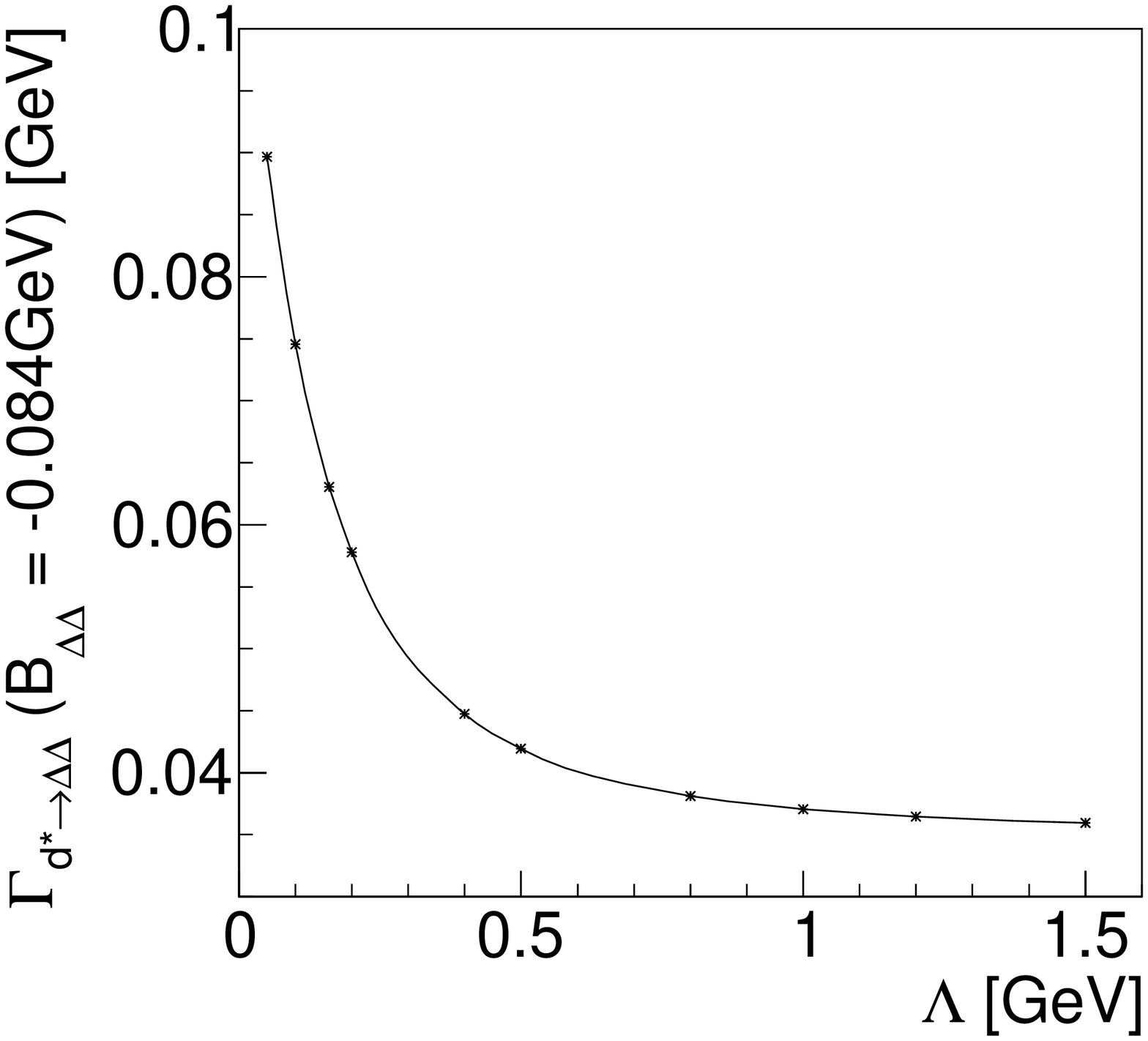}
        \includegraphics[width=0.29\textwidth,angle=0]{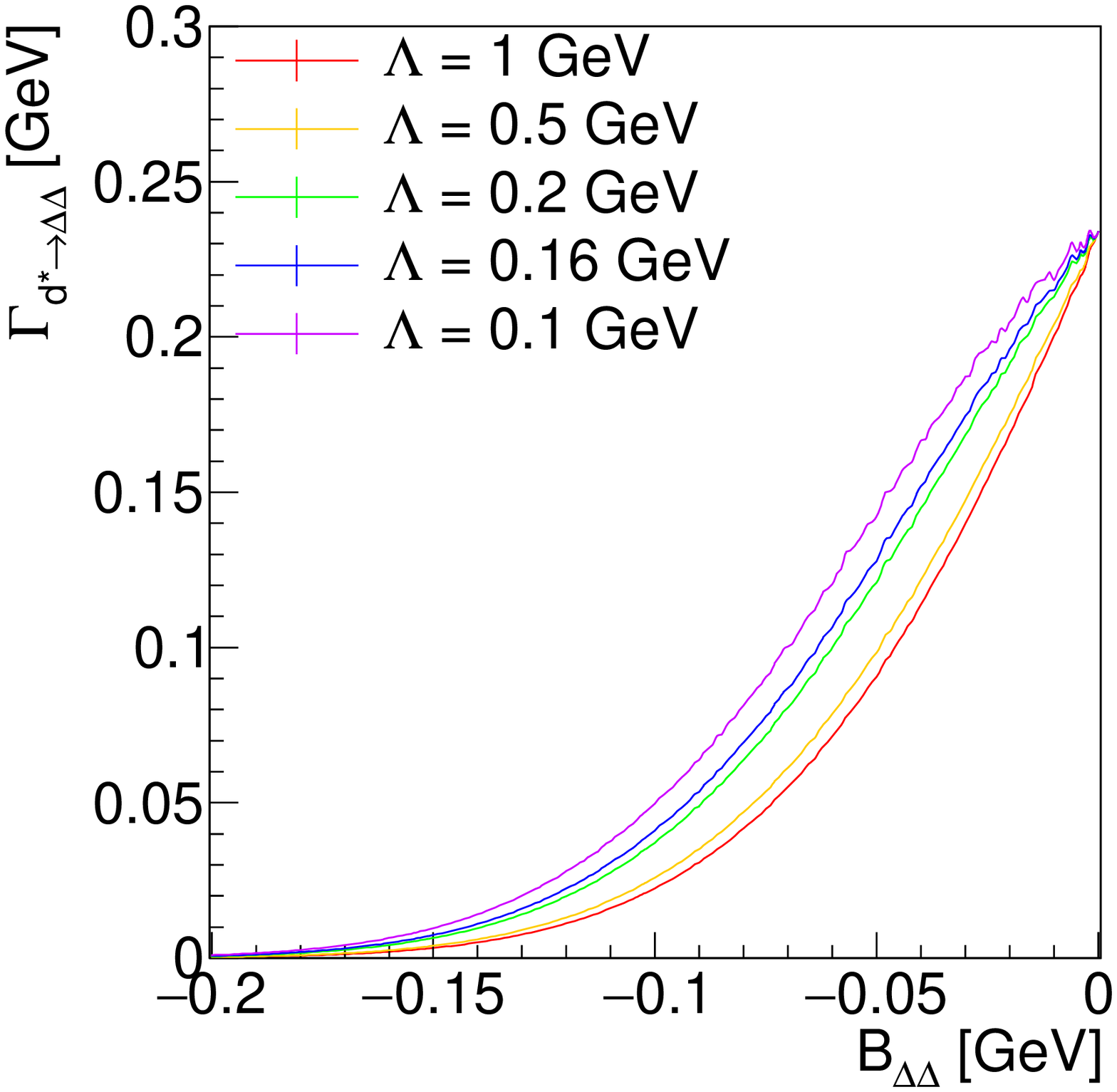}
\end{center}
\caption{$d^*(2380)$ width as a function of the Form Factor cut-off parameter $\Lambda$ (left) and partial width $\Gamma_{d^*\to\Delta\Delta}$ as a function of binding energy (right) for various $\Lambda$ values, $\Lambda = 0.1, 0.16, 0.2, 0.5, 1$~GeV/c in a rainbow order from violet to red.}
\label{d_width}
\end{figure}

We firstly explored the validity of the adopted cut off parameter $\Lambda =0.16$~GeV/c in the model.
The Form-Factor of the form Eq. 5. with a cut-off parameter $\Lambda =0.16$~GeV/c in the $d^*\to\Delta\Delta\to d\pi\pi$ was first introduced in a Ref.~\cite{MB} to explain the so-called ABC-effect, an enhancement in $M_{\pi\pi}$ close to threshold. Indeed, for the case where the nucleons in the deuteron have a very small relative momentum, there is a correspondence in the relative momentum between the $\Delta$'s and the relative momentum between the pions (and hence with the pion invariant mass). It was later speculated by A. Gal, that reduction of the $\Delta\Delta$ system size within the $d^*$  can lead to a further reduction of the $d^*$ width~\cite{Gal2}. To clarify the situation we have studied the predicted $d^*\to\Delta\Delta$ width dependence as a function of the cut of parameter $\Lambda$, Fig.~\ref{d_width}. The results indicate that the adopted $\Lambda =0.16$~GeV/c can not only reproduce the ABC effect, but also gives agreement with the measured $\Gamma(d^*\to\Delta\Delta)=62$~MeV. As evident in Fig~\ref{d_width}  we also reproduce the trend in which a higher value adopted for the cut-off parameter leads to a smaller width. 

\begin{figure}[!h]
\begin{center}
        \includegraphics[width=0.24\textwidth,angle=0]{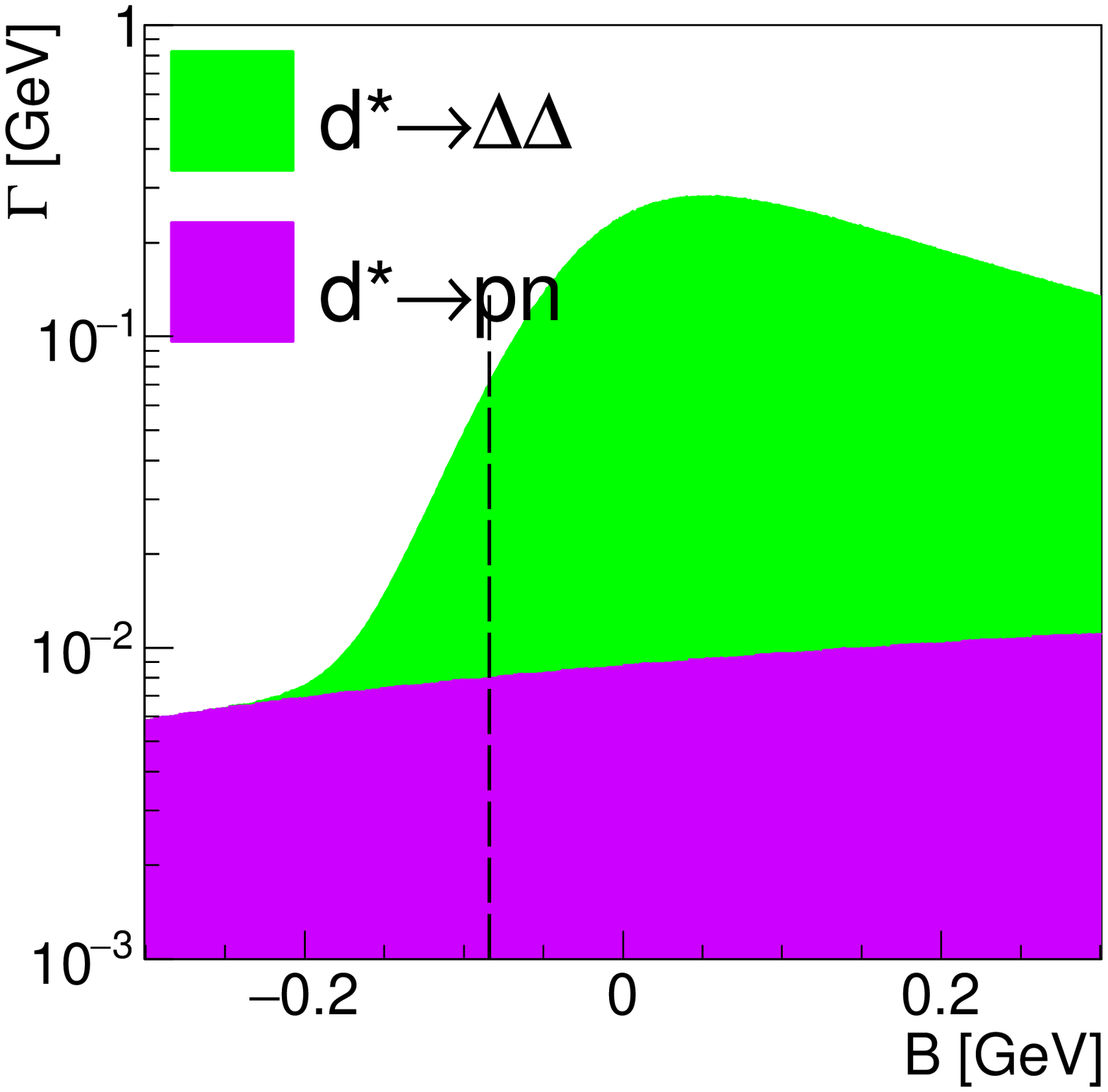}
        \includegraphics[width=0.24\textwidth,angle=0]{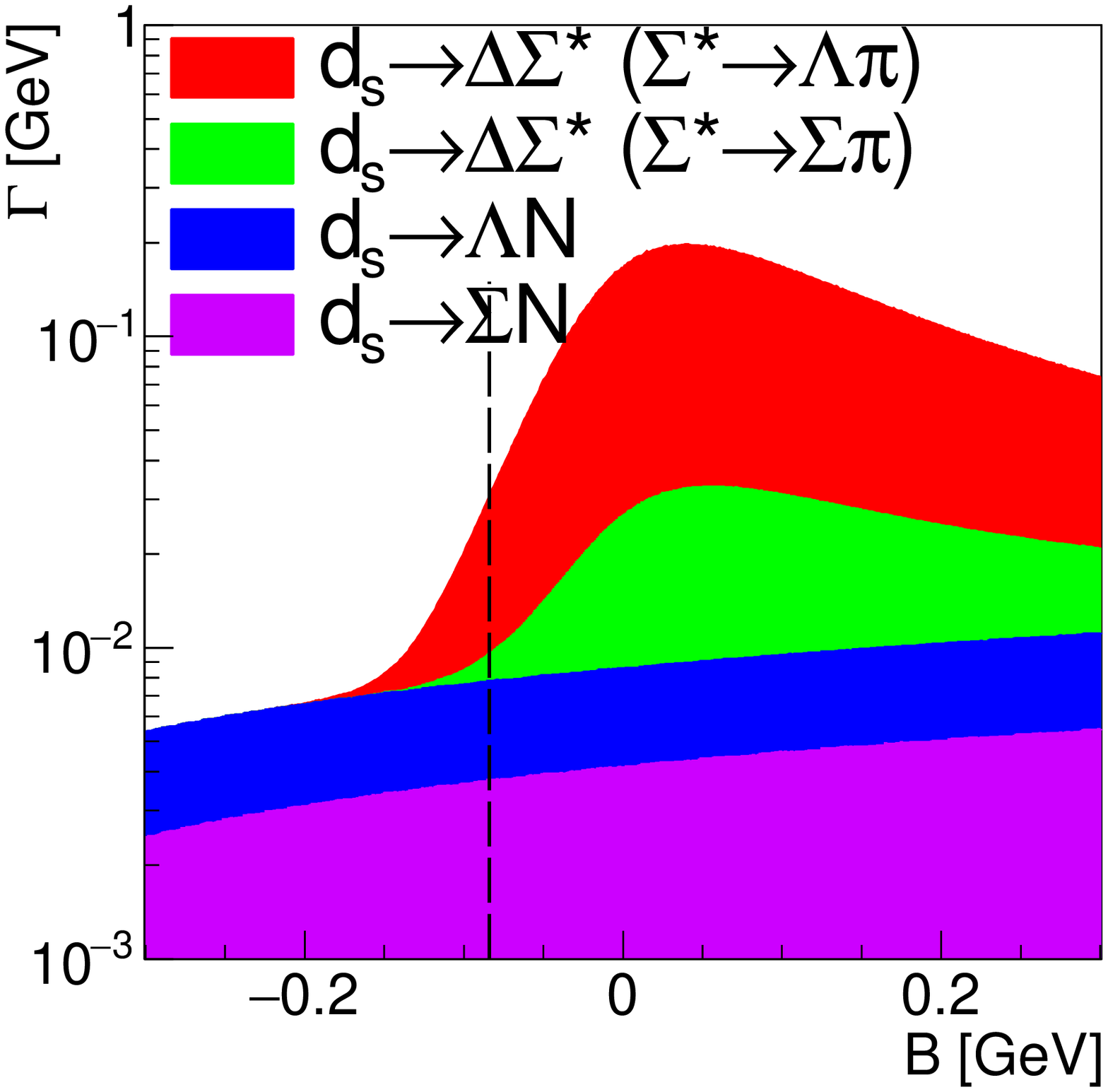}
        \includegraphics[width=0.24\textwidth,angle=0]{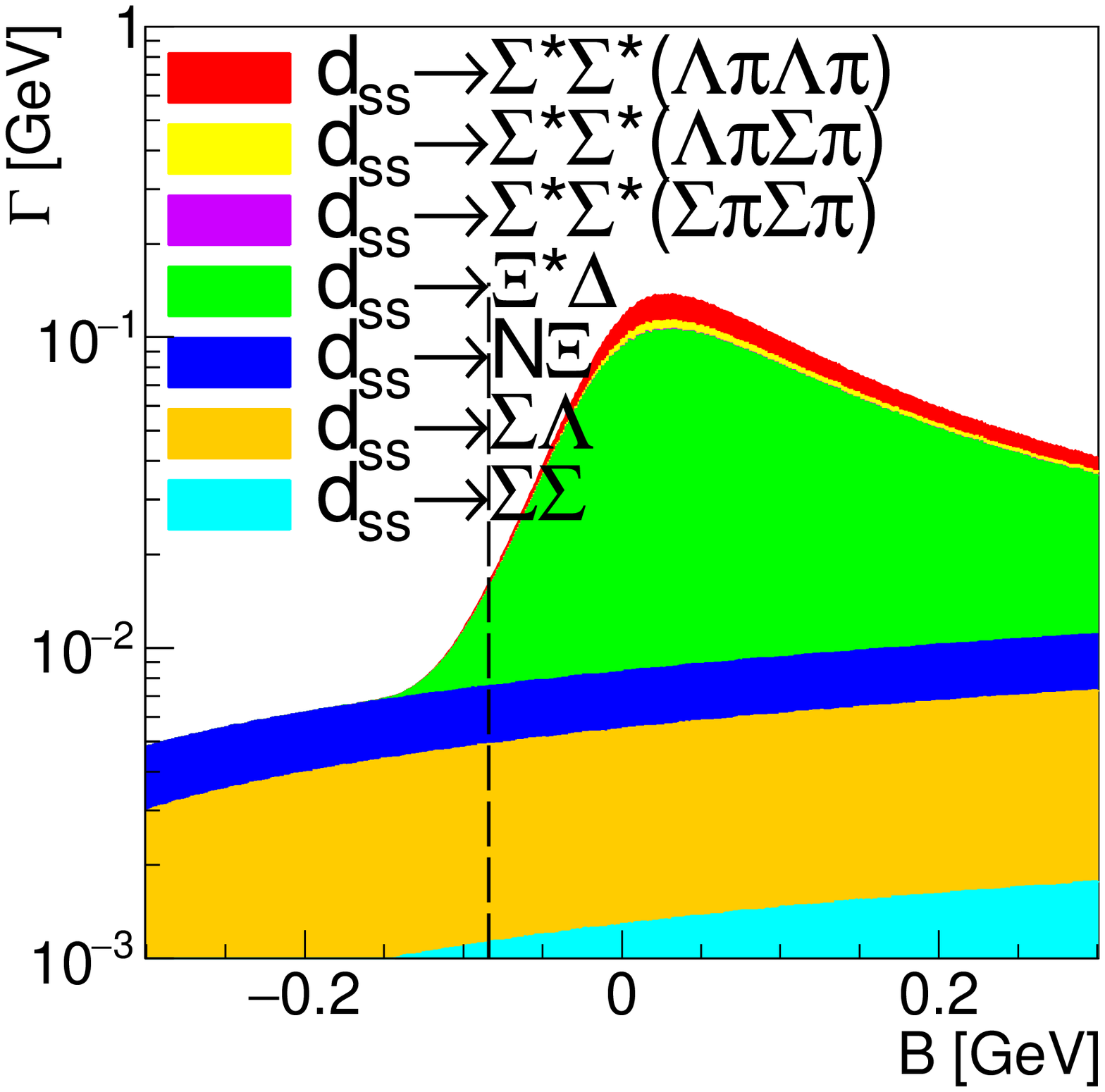}        \includegraphics[width=0.24\textwidth,angle=0]{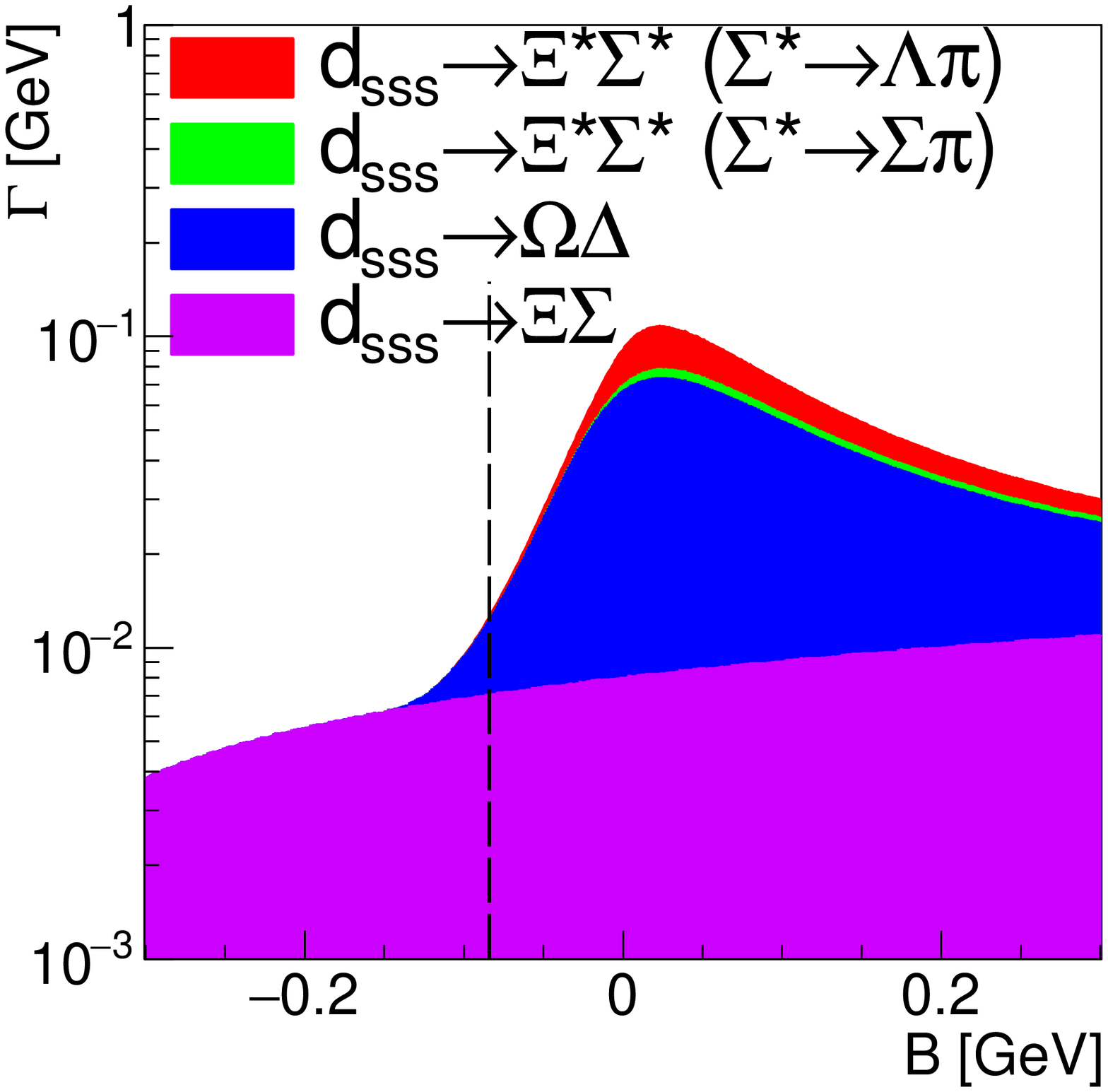}
\end{center}
\caption{$d^*$ multiplet total width as a function of binding energy (relative to the Decuplet-Decuplet pole) for the $d^*,d_{s},d_{ss},d_{sss}$ from left to right split into major decay branches (note the log scale). The vertical dashed line (common to all figures) shows the nominal expected mass, obtained under the assumption of the same binding for all multiplet members.}
\label{d_all_width}
\end{figure}

\begin{figure}[!h]
\begin{center}
        \includegraphics[width=0.24\textwidth,angle=0]{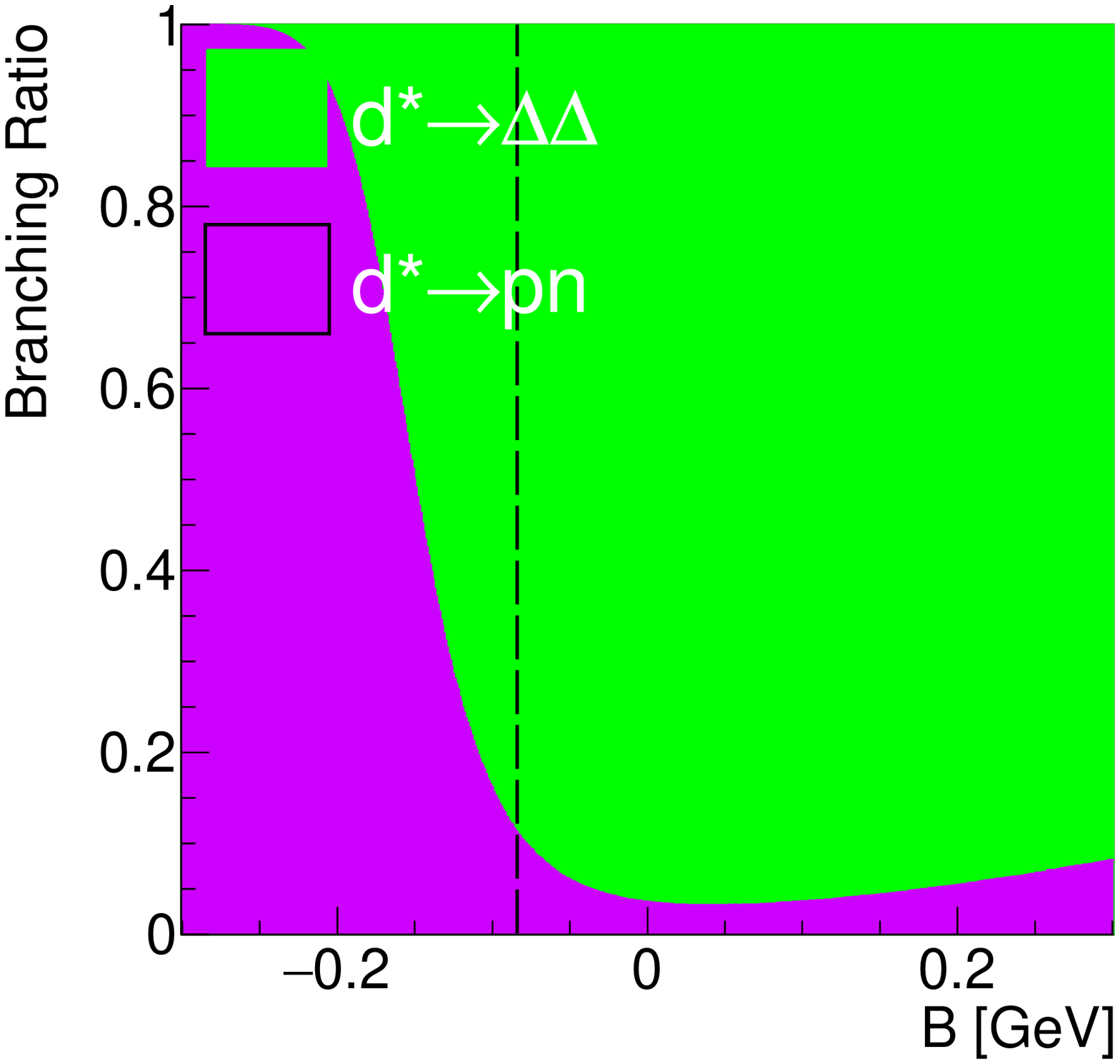}
        \includegraphics[width=0.24\textwidth,angle=0]{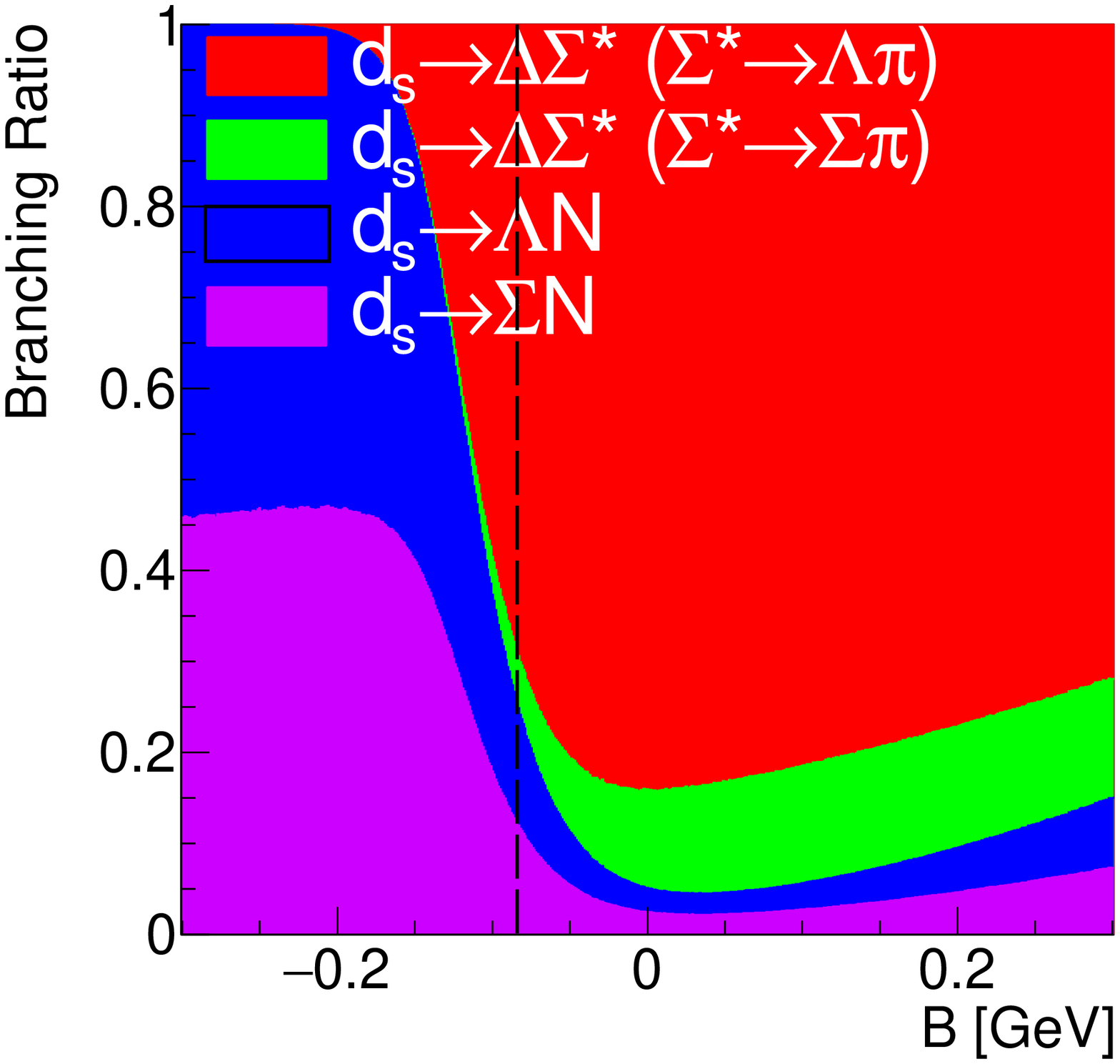}
        \includegraphics[width=0.24\textwidth,angle=0]{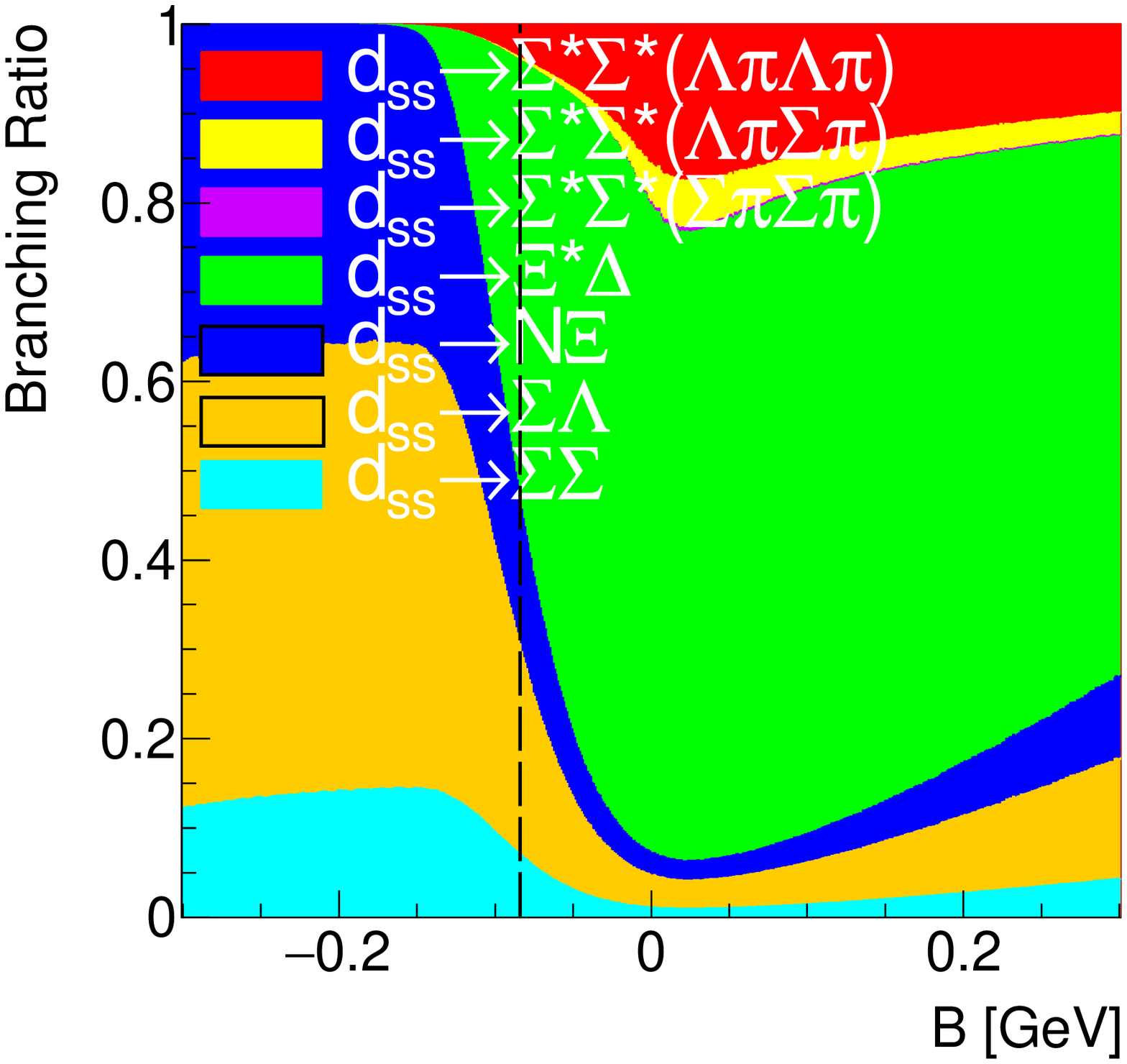}        \includegraphics[width=0.24\textwidth,angle=0]{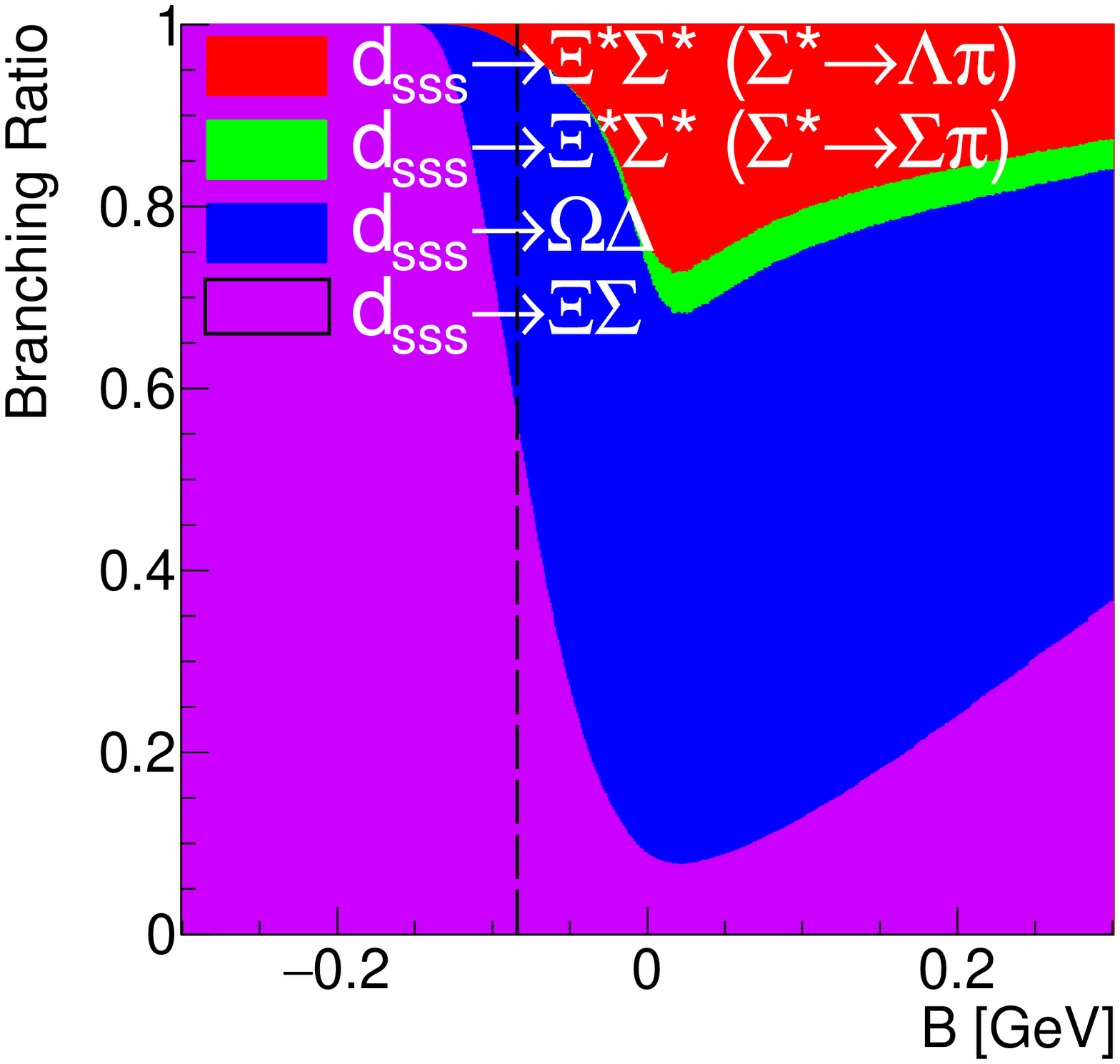}
\end{center}
\caption{$d^*$ multiplet branching ratio as a function of binding energy (relative to Decuplet-Decuplet pole) for the $d^*,d_{s},d_{ss},d_{sss}$ from left to right. The vertical dashed line (common to all figures) shows the nominal expected mass, obtained under the assumption of the same binding for all multiplet members.}
\label{d_all_Br}
\end{figure}

We therefore adopt $\Lambda =0.16$~GeV/c for subsequent calculations. The predicted width for all decuplet members is shown on Fig.~\ref{d_all_width}. In Fig.~\ref{d_all_Br} we show the predicted branching ratios as a function of binding energy. The results are summarised in Table.~\ref{Tab_width}

\begin{table}[ht]

\centering \protect\caption{$d^*$ multiplet width results for the $B= -84$~MeV. }
\vspace{2mm}
{%
\begin{tabular}{lcc|lcc|lcc|lcc}
\hline
\multicolumn{3}{|c|}{$d^*$} & \multicolumn{3}{|c|}{$d_s$} & \multicolumn{3}{|c|}{$d_{ss}$} & \multicolumn{3}{|c|}{$d_{sss}$} \\
\hline

decay & $\Gamma$, [MeV]  & BR [\%] & decay & $\Gamma$, [MeV]  & BR [\%] & decay & $\Gamma$, [MeV]  & BR [\%] & decay & $\Gamma$, [MeV]  & BR [\%]\\
$\Delta\Delta$ & 63 & 89 & $\Delta\Sigma^*(\Sigma^*\to\Lambda\pi)$ & 21 & 69 & $\Sigma^*\Sigma^*(\Lambda\Lambda\pi\pi)$ & 0.6 & 3.8 & $\Xi^*\Sigma^*(\Sigma^*\to\Lambda\pi)$ & 0.34 & 3 \\
$pn$ & 8 & 11 &  $\Delta\Sigma^*(\Sigma^*\to\Sigma\pi)$ & 1.7 & 6 & $\Sigma^*\Sigma^*(\Lambda\Sigma\pi\pi)$ & 0.05 & 0.3 & $\Xi^*\Sigma^*(\Sigma^*\to\Sigma\pi)$ & 0.0007 & 0.05 \\
  &  &  & $N\Lambda$ & 4.1 & 13 & $\Sigma^*\Sigma^*(\Sigma\Sigma\pi\pi)$ & 4.06e-6& 0 & $\Omega\Delta$ & 5.2 & 41  \\
  &  &  & $N\Sigma$ & 3.7 & 12 & $\Delta\Xi^*$ & 7.8 & 48.9 & $\Xi\Sigma$ & 7.1 & 56 \\
  &  &  &  & & & $N\Xi$ & 2.6 & 16.4 & & &  \\
  &  &  &  & & & $\Sigma\Sigma$ & 1.1 & 7.1 & & &  \\
  &  &  &  & & & $\Sigma\Lambda$ & 3.8 & 23.5 & & &  \\
  &  &  &  & & & & & & & &  \\
\hline
total & {\bf 71} &  & total  & {\bf 30.5} & & total & {\bf 15.95} & & total &  {\bf 12.6} &  \\
\hline
\end{tabular}} \label{Tab_width}
\end{table}

One can clearly see that the $8\oplus 8$ decays are predicted to be increasingly  important for the higher strangeness states, while both partial and total widths reduces substantially. For all hexaquark members only $10\oplus 10$ with $\Delta$ in a final state are important. For the $d_{ss}$ non-$\Delta$ channels ($\Sigma^*\Sigma^*\to$~anything) cover 4\% only. For the $d_{sss}$ the non-$\Delta$ ($\Xi^*\Sigma^*$) decay has 3\% branching ratio. It is interesting to note that the semi-electromagnetic decay branch $d_{sss}\to \Omega\Delta\to \Omega N\gamma$ is predicted to have a higher probability than the purely hadronic $d_{sss}\to \Xi^*\Sigma^* \to \Xi\Sigma\pi\pi$ decay. 

Hexaquarks are expected to be produced copiously in heavy ion collisions, however our estimations indicate the width for all $d^*$ multiplet states is rather large. Unfortunately this indicates their clean identification in the tough background conditions in typical heavy ion collisions may be challenging. The most feasible channel for such studies, having both a large partial width and a convenient isospin $3/2$ separation, is the $\Xi\Sigma$ branch in $d_{sss}$ decay. This could potentially be tested using $\Xi-\Sigma$ correlation functions.

For all channels the most prominent final states have high particle multiplicity, indicating hermetic detector apparati and exclusivity conditions would need to be established. The $d_{sss}$ is the only member with a dominant octet-octet decay, however the necessity of associated kaon production to conserve strangeness in the production mechanism also provides challenges.The requirements may be more easily achieved using beams containing intrinsic strangeness (e.g. Kaon beams as proposed in Ref.~\cite{KLF}).

\section{Summary}

We have developed a theoretical model, employing experimentally constrained parameters, to predict the possible decay branches and partial widths for all members of the $d^*$ hexaquark antidecuplet. For all strange-quark containing members of the antidecuplet the predicted widths are rather large, with the most promising decay channels including the broad $\Delta$ resonance in the final state. We demonstrated that a $d^*$ Form-Factor, which was first introduced to explain peculiarities in the $d^*\to d\pi\pi$ decay, can also explain the smallness of the $d^*$ width, in agreement  with the qualitative dimensional arguments of A. Gal. The results of the paper will be an important guide for the ongoing search for the $d^*$ anti-decuplet members employing photon-, pion-, and kaon-induced reactions, as well as in high energy collider experiments.

\begin{acknowledgements}
This work has been supported by the U.K. STFC ST/V002570/1, 	ST/P004008/1 and Strong2020 grants. 
\end{acknowledgements}


\begin{thebibliography}{9}
\bibitem {EndNote1} The appropriateness of the value for this cutoff parameter is explored in Ref~\cite{MB}, where it was shown to provided agreement with measured invariant mass distributions in the region of the so-called ABC-effect.
\end{thebibliography}
\end{document}